\documentclass[a4paper,12pt,titlepage]{article}

\usepackage{jtb}
\usepackage{amsmath,amssymb}
\usepackage{bm}
\usepackage[dvips]{graphicx}
\usepackage{color}
\usepackage{comment}
\usepackage{multirow}

\title{Indirect Reciprocity with Trinary Reputations}
\author{Shoma Tanabe$^{1}$, Hideyuki Suzuki$^{2}$, and Naoki Masuda$^{1,*}$\\
\\
$^1$ Department of Mathematical Informatics,\\
The University of Tokyo,\\
7-3-1 Hongo, Bunkyo, Tokyo 113-8656, Japan\\
\\
$^2$ Institute of Industrial Science,\\
The University of Tokyo,\\
4-6-1 Komaba, Meguro, Tokyo 153-8505, Japan\\
\\
$^*$ Corresponding author (masuda@mist.i.u-tokyo.ac.jp)\\
}
\date{\today}

\begin{document}
\bibliographystyle{jtb}
\maketitle
\begin{abstract}
Indirect reciprocity is a reputation-based mechanism for cooperation in social dilemma situations when individuals do not repeatedly meet.
The conditions under which cooperation based on indirect reciprocity occurs have been examined in great details.
Most previous theoretical analysis assumed for mathematical tractability that an individual possesses a binary reputation value, i.e., good or bad, which depends on their past actions and other factors.
However, in real situations, reputations of individuals may be multiple valued.
Another puzzling discrepancy between the theory and experiments is the status of the so-called image scoring, in which cooperation and defection are judged to be good and bad, respectively, independent of other factors.
Such an assessment rule is found in behavioral experiments, whereas it is known to be unstable in theory.
In the present study, we fill both gaps by analyzing a trinary reputation model.
By an exhaustive search, we identify all the cooperative and stable equilibria composed of a homogeneous population or a heterogeneous population containing two types of players.
Some results derived for the trinary reputation model are direct extensions of those for the binary model.
However, we find that the trinary model allows cooperation under image scoring under some mild conditions.
\end{abstract}
\section{Introduction}
Humans and other animals often cooperate even when cooperation is more costly than defection.
In such social dilemma situations, direct reciprocity is among main reasons for cooperation between pairs of individuals that repeatedly meet each other \cite{Trivers1971QRB,Axelrod1984book}.
However, individuals, in particular humans, cooperate with others even when they seldom meet the same partners more than once, as is the case for large populations.
Reputation-based indirect reciprocity (also called downstream reciprocity; we simply call it indirect reciprocity in this paper) seems to be a dominant mechanism that enables cooperation in this situation.
In indirect reciprocity, individuals cooperate with others with good reputations and they in turn gain good reputations if they appositely behave (e.g., cooperate) toward somebody else.
The conditions under which indirect reciprocity realizes cooperation have been theoretically and numerically clarified in great details \cite{Nowak1998aJTB,Nowak1998bNature,Leimar2001PRSB,Panchanathan2003JTB,Mohtashemi2003JTB,Fishman2003JTB,Ohtsuki2004JTB,OhtsukiIwasa2004JTB,Ohtsuki2006JTB,Ohtsuki2007JTB,Nowak2005Nature,Brandt2004JTB,Brandt2005PNAS,Brandt2006JTB,Pacheco2006PLoS,Roberts2008PRSB,Uchida2010PRE,Nakamura2011PLos,Berger2011GEB,Sigmund2012JTB}.

Under the so-called image scoring, cooperation and defection are regarded to be good and bad behavior, respectively \cite{Nowak1998aJTB,Nowak1998bNature}.
Laboratory experiments suggest that humans use image scoring to evaluate others' behavior \cite{Wedekind2000Science,Milinski2001PRSB,Seinen2006EER}.
However, main theories attain that image scoring does not stabilize cooperation \cite{Leimar2001PRSB,Panchanathan2003JTB,Ohtsuki2004JTB,OhtsukiIwasa2004JTB,Ohtsuki2007JTB,Roberts2008PRSB}.
Although some studies have shown the viability of cooperation under image scoring \cite{Nowak1998aJTB,Fishman2003JTB,Brandt2004JTB,Brandt2005PNAS,Brandt2006JTB,Uchida2010PRE}, the situations in which cooperation occurs are, in our view, quite restricted (but see \cite{Berger2011GEB}; we discuss this reference in Discussion).
Under image scoring, cooperation occurs only when individuals always cooperate in the first round \cite{Nowak1998aJTB}, unconditional defectors sometimes cooperate \cite{Fishman2003JTB}, the number of interaction obeys the binomial distribution \cite{Brandt2004JTB} or Poisson distribution \cite{Brandt2006JTB}, the probability that individuals recognize others' reputations increases in time \cite{Brandt2005PNAS}, or the reputations of individuals are revealed to others with a small probability \cite{Uchida2010PRE}.
Therefore, the reason for the discrepancy between the experiments and theory remains obscure.

For mathematical tractability and possible influences of the first seminal theoretical papers on this subject \cite{Nowak1998aJTB,Nowak1998bNature}, most theoretical results of indirect reciprocity are derived from the analysis of binary reputation models.
In other words, individuals are endowed with the binary reputation, i.e., good ($+$) or bad ($-$), depending on the last action toward others and other factors.
However, the binary reputation may not be realistic in that only the last behavior of an individual in the social dilemma situation determines the reputation of the individual.
In fact, experimental \cite{Wedekind2000Science,Milinski2001PRSB,Seinen2006EER,Milinski2002Nature,Wedekind2002CB,Keser2003IBMSJ,Bolton2004MS,Bolton2005JPE,Engelmann2009GEB} and numerical \cite{Diekmann2005ESSA} studies of indirect reciprocity in the context of online marketplaces often assume that the reputations are many valued, which complies with the reality of online marketplaces \cite{Resnick2002AAM,Resnick2006ECON}.
More than binary valued reputations have also been employed in numerical studies of indirect reciprocity in theoretical biology literature \cite{Nowak1998bNature,Leimar2001PRSB,Mohtashemi2003JTB,Roberts2008PRSB}.
Nevertheless, these studies are not concerned with relationships between the degree of cooperation and the number of the possible reputation values.

In this paper we analyze a trinary reputation model to identify stable populations that realize cooperation.
The difference between the present results and those derived from the binary reputation models is remarkable.
In particular, we find that image scoring can stabilize cooperation in the trinary reputation model.
\section{Model}
\subsection{Donation Game with Reputations}\label{sec:donation}
\begin{figure}[htbp]
 \begin{center}
  \includegraphics[width=\linewidth]{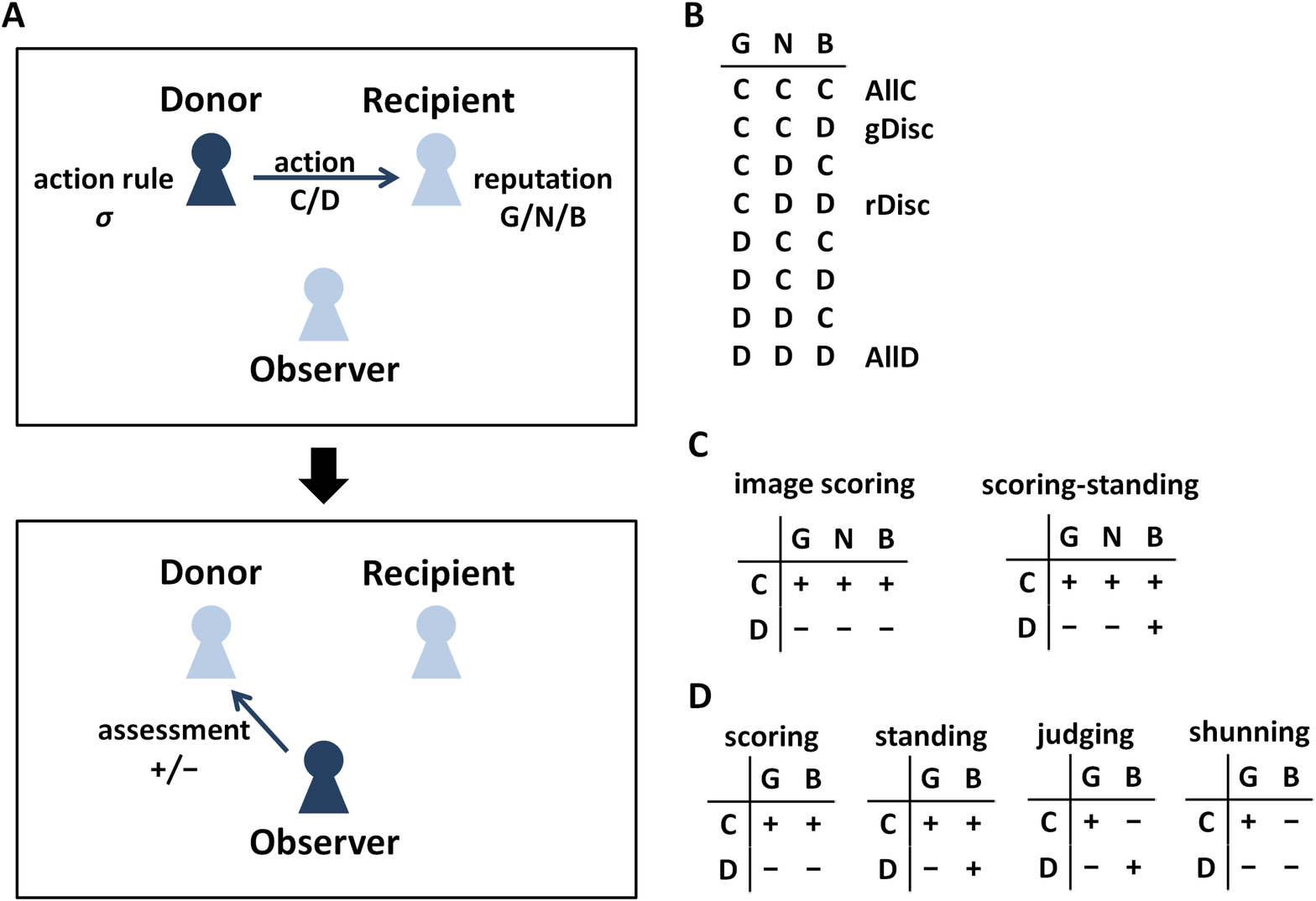}
 \end{center}
 \caption{
Rule of the donation game with trinary reputations.
(A) Illustration of the interaction in a single game.
(B) Eight action rules.
(C) Representative social norms.
The rows represent the donor's actions (i.e., C and D), the columns represent the recipient's reputations (${\rm G}$, ${\rm N}$, and ${\rm B}$), and $+$ and $-$ represent the assessments that observer assigns to the donor.
(D) Representative social norms in the binary reputation model.
}
 \label{fig:model}
\end{figure}

We consider an infinitely large population.
In each generation, the so-called donation game is repeated for sufficiently many rounds.
Figure~\ref{fig:model}(A) illustrates the interaction in each round.
Two players are randomly selected from the population, one as donor and the other as recipient, with the equal probability.
The donor intends to cooperate (C) or defect (D) toward the recipient according to the action rule $\sigma$, which we define below.
We assume that the donor misimplements intended C such that the donor actually defects with probability $\epsilon_{\rm i}>0$ and that the intended D is always correctly implemented.
We seek the possibility of cooperation in the population under this kind of implementation error, which is adverse to cooperation \cite{Panchanathan2003JTB,Fishman2003JTB,Ohtsuki2004JTB,OhtsukiIwasa2004JTB,Ohtsuki2006JTB,Ohtsuki2007JTB,Nowak2005Nature,Brandt2004JTB,Brandt2005PNAS,Brandt2006JTB,Uchida2010PRE,Berger2011GEB}.
If the donor implements C, the donor pays cost $c$, and the recipient obtains benefit $b$.
If the donor implements D, the payoffs to the donor and recipient do not change.
We assume that $0<c<b$ such that the donation game is essentially the prisoner's dilemma.

We assume that each player possesses a reputation that takes one of the three values, i.e., ${\rm G}$ (Good), ${\rm N}$ (Neutral), or ${\rm B}$ (Bad).
The action rule $\sigma$ is a function from the recipient's reputation to the donor's intended action (i.e., C or D).
Therefore, there are $2^3=8$ action rules, as shown in Fig.~\ref{fig:model}(B).
For example, the AllC and AllD intend C and D regardless of the recipient's reputation, respectively.
The so-called generous discriminator (gDisc) intends C when the recipient's reputation is either ${\rm G}$ or ${\rm N}$ and D otherwise.
The so-called rigorous discriminator (rDisc) intends C when the recipient's reputation is ${\rm G}$ and D otherwise.

At the end of each round, the observer assesses the donor to be $+$ or $-$ depending on the donor's implemented action (i.e., C or D) and the recipient's reputation (i.e., ${\rm G}$, ${\rm N}$, or ${\rm B}$).
Such an assessment rule is a function from the donor's implemented action and the recipient's reputation to the observer's assessment and is called the second-order social norm, which extends the concept in the case of the binary reputation \cite{Leimar2001PRSB,Nowak2005Nature}.
There are $2^{2\times3}=64$ social norms.
For example, under the social norm called image scoring (Fig.~\ref{fig:model}(C)), the observer assigns $+$ and $-$ to the donor's actions C and D, respectively, regardless of the recipient's reputation.
Another social norm named scoring--standing is also shown in Fig.~\ref{fig:model}(C).
Finally, we assume that the assessment (i.e., $+$ or $-$) that the donor receives is opposite to the one assigned by the observer with probability $\epsilon_{\rm a}>0$ \cite{OhtsukiIwasa2004JTB,Ohtsuki2007JTB,Nakamura2011PLos}.
We assume that the donor's reputation is publicly shared in the population.
In other words, the donor's new reputation is instantaneously known to all the players in the population.

We repeat many rounds of the game and define the payoff to each player as the sum of the payoff obtained in the games that involve the player.
Because the population is infinite, each pair of players plays the game at most once such that direct reciprocity is excluded.

For a later use, we mention four representative second-order social norms in the binary reputation model (Fig.~\ref{fig:model}(D)).
Image scoring  (``scoring'' in Fig.~\ref{fig:model}(D)) does not stabilize cooperation \cite{Panchanathan2003JTB,Ohtsuki2004JTB,OhtsukiIwasa2004JTB,Ohtsuki2007JTB,Roberts2008PRSB} unless somewhat strong conditions are met \cite{Nowak1998aJTB,Fishman2003JTB,Brandt2004JTB,Brandt2005PNAS,Brandt2006JTB,Uchida2010PRE} (but see \cite{Berger2011GEB}).
Simple standing (``standing'' in Fig.~\ref{fig:model}(D)) and stern judging (``judging'' in Fig.~\ref{fig:model}(D); also called Kandori \cite{Kandori1992RES}) are known to stabilize cooperation \cite{OhtsukiIwasa2004JTB}.
Shunning stabilizes cooperation under certain conditions \cite{Ohtsuki2007JTB,Nakamura2011PLos}.
\subsection{Reputation Dynamics}\label{sec:dynamics}
\begin{figure}[htbp]
  \begin{center}
  \includegraphics[width=0.5\linewidth]{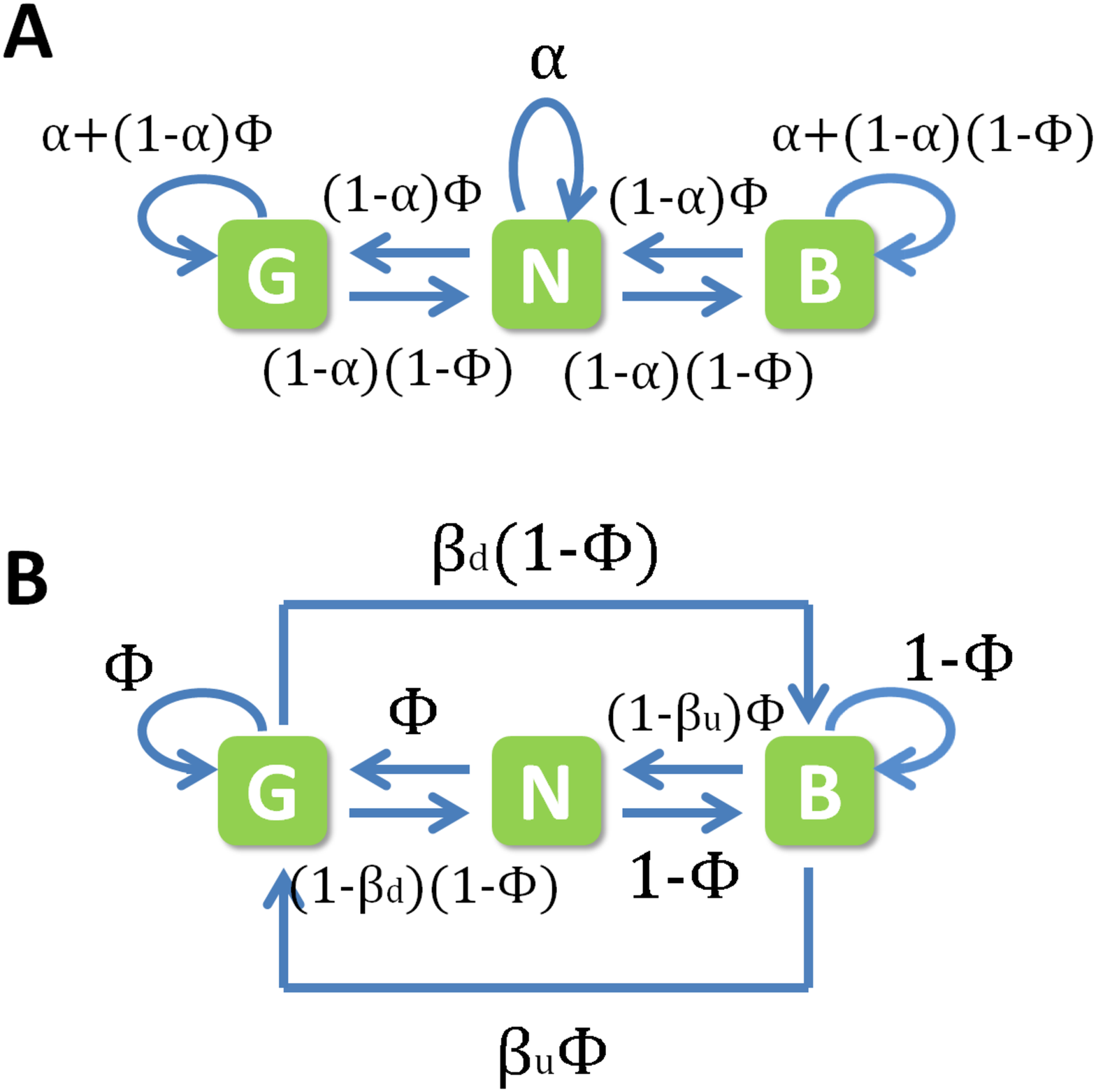}
  \end{center}
 \caption{
Two types of reputation dynamics.
(A) Gradual reputation dynamics. 
(B) Saltatory reputation dynamics.
$\Phi$ represents the probability that the donor receives $+$.
}
 \label{fig:dynamics}
\end{figure}

After each round, the donor's reputation is updated on the basis of the observer's assessment.
Basically, the donor's reputation shifts upward and downward if the donor receives $+$ and $-$, respectively.
When the reputation is binary (i.e., ${\rm G}$ and ${\rm B}$), ${\rm G}$ and ${\rm B}$ are equivalent to $+$ and $-$ that the donor receives in the last game, respectively \cite{Nowak1998aJTB,Nowak2005Nature}.
However, the relationship between the reputation and assessment is not straightforward in the trinary reputation model.
We assume that reputation dynamics obey a Markov chain.
We consider the reputation dynamics illustrated in Fig.~\ref{fig:dynamics}.

In the reputation dynamics shown in Fig.~\ref{fig:dynamics}(A), which we call the gradual dynamics, the reputation is assumed to move by at most one level in each round.
The reputation is unchanged with probability $\alpha$ ($0\le \alpha < 1$).

In the reputation dynamics shown in Fig.~\ref{fig:dynamics}(B), which we call the saltatory dynamics, the donor's reputation can transit from ${\rm G}$ to ${\rm B}$ or vice versa in one step.
When a ${\rm G}$ donor receives $-$, the donor's new reputation becomes ${\rm B}$ and ${\rm N}$ with probabilities $\beta_{\rm d}$ and $1-\beta_{\rm d}$, respectively $(0 \le \beta_{\rm d} \le 1)$.
When a ${\rm B}$ donor receives $+$, the donor's new reputation becomes ${\rm G}$ and ${\rm N}$ with probabilities $\beta_{\rm u}$ and $1-\beta_{\rm u}$, respectively $(0 \le \beta_{\rm u} \le 1)$.
When $(\beta_{\rm d},\beta_{\rm u})=(1,0)$, the reputation dynamics are similar to the so-called $T$-period punishment with $T=2$ \cite{Kandori1992RES} in which players have either state $0$ (innocent), $1$ (guilty and no repent), ..., or $T$ (guilty and $T-1$ times of repentant behavior).
Guilty players in state $1$ regain the innocent state by cooperating with innocent players successive $T$ times.
Therefore, the states $0$, $1$, and $2$ are similar to reputations ${\rm G}$, ${\rm B}$, and ${\rm N}$, respectively, in our model.
When $(\beta_{\rm d},\beta_{\rm u})=(0,1)$, the reputation dynamics are similar to the so-called tolerant scoring \cite{Berger2011GEB} because the donor's reputation becomes ${\rm B}$ if and only if the donor receives $-$ in the last two rounds, and the donor obtains a ${\rm G}$ reputation if the donor cooperates just once.
\section{Analysis Methods}
\subsection{Homogeneous Populations}\label{sec:homo}
We first examine the stability of a homogeneous population of resident players with action rule $\sigma$ against mutants with different action rules under a given social norm.
Let $p_{\rm G}$, $p_{\rm N}$, and $p_{\rm B}$ be the probabilities that the reputation of a resident player is ${\rm G}$, ${\rm N}$, and ${\rm B}$, respectively.
After a transient of the reputation dynamics, the three probabilities converge to the equilibrium values denoted by $p_{\rm G}^*$, $p_{\rm N}^*$, and $p_{\rm B}^*$.
For expository purposes, we focus on the gradual reputation dynamics (Fig.~\ref{fig:dynamics}(A)) in this section.
The following calculations are similar for the saltatory reputation dynamics (Fig.~\ref{fig:dynamics}(B)); the corresponding results are shown in Appendix A.
For the gradual reputation dynamics, we obtain
\begin{eqnarray}
 \left \{
 \begin{array}{rcl}
  p_{\rm G}^*&=&p_{\rm G}^* [\alpha+(1-\alpha)\Phi^*] + p_{\rm N}^* (1-\alpha)\Phi^*,\vspace{-2mm}\cr
  p_{\rm N}^*&=&p_{\rm G}^* (1-\alpha)(1-\Phi^*) + p_{\rm N}^* \alpha + p_{\rm B}^* (1-\alpha) \Phi^*,\vspace{-2mm}\cr
  p_{\rm B}^*&=&p_{\rm N}^* (1-\alpha)(1-\Phi^*) + p_{\rm B}^* [\alpha+(1-\alpha)(1-\Phi^*)],
 \end{array}
 \right .
\label{eq:p^star}
\end{eqnarray}
where $\Phi^*$ is the probability that the donor receives $+$ in the equilibrium.
Equation~\eqref{eq:p^star} and the normalization $p_{\rm G}^*+p_{\rm N}^*+p_{\rm B}^*=1$ lead to
\begin{equation}
 (p_{\rm G}^*,p_{\rm N}^*,p_{\rm B}^*)=\left(\displaystyle \frac{{\Phi^*}^2}{1-\Phi^*+{\Phi^*}^2},\frac{\Phi^*(1-\Phi^*)}{1-\Phi^*+{\Phi^*}^2},\frac{(1-\Phi^*)^2}{1-\Phi^*+{\Phi^*}^2}\right).
 \label{eq:p^star2}
\end{equation}
It should be noted that Eq.~\eqref{eq:p^star2} and the following results are independent of the value of $\alpha$.
Equation~\eqref{eq:p^star2} implies that the distribution of the reputation is monotonous in the sense that $p_{\rm G}^*>p_{\rm N}^*>p_{\rm B}^*$ if $\Phi^*>0.5$ and $p_{\rm G}^*<p_{\rm N}^*<p_{\rm B}^*$ if $\Phi^*<0.5$.

$\Phi^*$ is given by
\begin{equation}
 \Phi^*=\sum_{r \in \{\rm{G,N,B}\}} p_r^* [\zeta_r \Phi_{{\rm C},r} + (1-\zeta_r) \Phi_{{\rm D},r}],
 \label{eq:Phi^star}
\end{equation}
where $\zeta_r$ represents the probability that the donor's implemented action is C when the recipient has reputation $r \in \{\rm{G,N,B}\}$.
$\zeta_r$ has a one-to-one correspondence with action rule $\sigma$.
For example, AllC, gDisc, rDisc, and AllD are equivalent to $(\zeta_{\rm G},\zeta_{\rm N},\zeta_{\rm B})=(1-\epsilon_{\rm i},1-\epsilon_{\rm i},1-\epsilon_{\rm i})$, $(1-\epsilon_{\rm i},1-\epsilon_{\rm i},0)$, $(1-\epsilon_{\rm i},0,0)$, and $(0,0,0)$, respectively.
$\Phi_{{\rm C},r}$ and $\Phi_{{\rm D},r}$ in Eq.~\eqref{eq:Phi^star} are the probabilities that the donor receives $+$ when the donor's action is C and D, respectively, and the recipient has reputation $r \in \{\rm{G,N,B}\}$.
For example, under image scoring (Fig.~\ref{fig:model}(C)), $\Phi_{{\rm C},r}=1-\epsilon_{\rm a}$ and $\Phi_{{\rm D},r}=\epsilon_{\rm a}$ for any $r$.
Under the so-called scoring--standing (see Sec.~\ref{sec:gradual} for the notation of the social norms) shown in Fig.~\ref{fig:model}(C), $\Phi_{{\rm C},r}=1-\epsilon_{\rm a}$ for any $r$, $\Phi_{{\rm D},+}=\Phi_{{\rm D},0}=\epsilon_{\rm a}$, and  $\Phi_{{\rm D},-}=1-\epsilon_{\rm a}$.
Each term on the right-hand side of Eq.~\eqref{eq:Phi^star} is a multiplication of three probabilities, i.e., 
(i) probability $p_r^*$ that a recipient with reputation $r$ is selected, 
(ii) probability $\zeta_r$ or $1-\zeta_r$ that a donor implements C or D, respectively, 
and (iii) probability $\Phi_{{\rm C},r}$ or $\Phi_{{\rm D},r}$ that the observer assigns $+$ to the donor.

We substitute $p_{\rm G}^*$, $p_{\rm N}^*$, and $p_{\rm B}^*$ obtained from Eq.~\eqref{eq:p^star2} in the right-hand side of Eq.~\eqref{eq:Phi^star}, which we denote by $f(\Phi^*)$.
We obtain $\Phi^*$ by solving $x=f(x)$, $0 \le x \le 1$.
Because $f(x)$ has a quadratic numerator and denominator in terms of $x$, equation $x=f(x)$ has at most three solutions.
Under any pair of action rule and social norm, to which we refer as action--norm pair in the following, $0<\epsilon_{\rm a} \le f(x) \le (1-\epsilon_{\rm i})(1-\epsilon_{\rm a}) <1$ holds true for any $0 \le x \le 1$.
Therefore, the iteration scheme, in which we start with an initial $x$ value ($0 \le x \le 1$) and repeatedly apply $f$, always converges.
If the iteration starting from $x=0$ and that starting from $x=1$ converge to the same value, the solution specified by $\Phi^*$ and $\{ p_{\rm G}^*,p_{\rm N}^*,p_{\rm B}^*\}$ is unique.
We confirmed that $x=f(x)$ has a unique solution for each of the $8 \times 64=512$ action--norm pairs.

Let $\Psi(\sigma,\{p_{\rm G},p_{\rm N},p_{\rm B}\})$ be the probability that a donor with action rule $\sigma$ cooperates with a recipient randomly chosen from a population according to reputation distribution $\{p_{\rm G},p_{\rm N},p_{\rm B}\}$.
We obtain
\begin{equation}
 \Psi(\sigma,\{p_{\rm G},p_{\rm N},p_{\rm B}\})=p_{\rm G} \zeta_{\rm G} + p_{\rm N} \zeta_{\rm N} + p_{\rm B} \zeta_{\rm B}.
 \label{eq:Psi}
\end{equation}
The average payoff per round to a resident player in the homogeneous population is given by
\begin{equation}
 \pi=-c \Psi(\sigma,\{p_{\rm G}^*,p_{\rm N}^*,p_{\rm B}^*\})+b \Psi(\sigma,\{p_{\rm G}^*,p_{\rm N}^*,p_{\rm B}^*\}).
 \label{eq:pi}
\end{equation}

To examine the stability of the homogeneous population, we consider an infinitesimally small fraction of mutants with action rule $\sigma^\prime (\ne \sigma)$ that invades the homogeneous resident population.
The equilibrium probability distribution of the mutant's reputation, denoted by $\{p_{\rm G}^{\prime *},p_{\rm N}^{\prime *},p_{\rm B}^{\prime *}\}$,
the equilibrium probability that a mutant receives $+$, denoted by $\Phi^{\prime *}$, and
payoff to a mutant player, denoted by $\pi^\prime$, are given by
\begin{equation}
 (p_{\rm G}^{\prime *},p_{\rm N}^{\prime *},p_{\rm B}^{\prime *})=\left(\displaystyle \frac{{\Phi^{\prime *}}^2}{1-\Phi^{\prime *}+{\Phi^{\prime *}}^2},\frac{\Phi^{\prime *}(1-\Phi^{\prime *})}{1-\Phi^{\prime *}+{\Phi^{\prime *}}^2},\frac{(1-\Phi^{\prime *})^2}{1-\Phi^{\prime *}+{\Phi^{\prime *}}^2}\right),
 \label{eq:p^star2_mut}
\end{equation}
\begin{equation}
 \Phi^{\prime *}=\sum_{r \in \{\rm{G,N,B}\}} p_r^* [\zeta^\prime_r \Phi_{{\rm C},r} + (1-\zeta^\prime_r) \Phi_{{\rm D},r}],
 \label{eq:Phi^star_mut}
\end{equation}
and
\begin{equation}
 \pi^\prime=-c \Psi(\sigma^\prime,\{p_{\rm G}^*,p_{\rm N}^*,p_{\rm B}^*\})+b \Psi(\sigma,\{p_{\rm G}^{\prime *},p_{\rm N}^{\prime *},p_{\rm B}^{\prime *}\}),
 \label{eq:pi_mut}
\end{equation}
respectively.
In Eq.~\eqref{eq:Phi^star_mut}, $\zeta^\prime_r$ represents the probability that a mutant cooperates with a recipient with reputation $r$.
It should be noted that $p_r^*$ in Eq.~\eqref{eq:Phi^star_mut} is the solution of Eqs.~\eqref{eq:p^star2} and \eqref{eq:Phi^star} and that $\Phi_{{\rm C},r}$ and $\Phi_{{\rm D},r}$ also refer to the values for the resident population.

Action rule $\sigma$ adopted by the resident players is strict Nash equilibrium if the payoff to a resident player (i.e., $\pi$) is larger than the payoff to any mutant player (i.e., $\pi^\prime$).
Using Eqs.~\eqref{eq:pi} and \eqref{eq:pi_mut}, we obtain this condition as follows:
\begin{equation}
 \begin{split}
  \displaystyle &\frac{b}{c} \left[\Psi(\sigma,\{p_{\rm G}^{\prime *},p_{\rm N}^{\prime *},p_{\rm B}^{\prime *}\})- \Psi(\sigma,\{p_{\rm G}^*,p_{\rm N}^*,p_{\rm B}^*\})\right]\cr
               <& \Psi(\sigma^\prime,\{p_{\rm G}^*,p_{\rm N}^*,p_{\rm B}^*\}) - \Psi(\sigma,\{p_{\rm G}^*,p_{\rm N}^*,p_{\rm B}^*\}) \ \ \ {\rm for\ any\ }\sigma^\prime \ne \sigma.
 \end{split}
 \label{eq:NE}
\end{equation}

We also investigate the stability of action rule $\sigma$ against invasion by a previously identified strong competitor \cite{Leimar2001PRSB}, which is so-called the Self strategy \cite{OhtsukiIwasa2004JTB}.
A Self donor plays a donation game as selfishly as possible under the constraint that the donor's reputation stays above a threshold.
When determining the action, the Self donor refers to its own reputation and does not refer to the recipient's reputation.
We assume that the Self donor defects if its reputation is G and cooperates otherwise.
Under the gradual reputation dynamics, the Self player maintains its reputation value at G or N, not B, except in the case of error.
\subsection{Heterogeneous Populations Composed of Two Action Rules}\label{sec:hetero}
We also examine the stability of heterogeneous populations in which two action rules, denoted by $\sigma_1$ and $\sigma_2$, coexist with fractions $q_1$ and $q_2$, respectively ($q_1+q_2=1$).
For each of $(q_1,q_2)=(0.01,0.99)$, $(0.02,0.98)$, ..., $(0.99,0.01)$), we first calculate the equilibrium probabilities of $+$ for $\sigma_1$ and $\sigma_2$ by an iteration scheme similar to that described in Sec.~\ref{sec:homo}.
Under the gradual reputation dynamics, the distribution of the reputation values is given by
\begin{equation}
 (p_{{\rm G},i}^*,p_{{\rm N},i}^*,p_{{\rm B},i}^*)=\left(\displaystyle \frac{{\Phi_i^*}^2}{1-\Phi_i^*+{\Phi_i^*}^2},\frac{\Phi_i^*(1-\Phi_i^*)}{1-\Phi_i^*+{\Phi_i^*}^2},\frac{(1-\Phi_i^*)^2}{1-\Phi_i^*+{\Phi_i^*}^2}\right),
 \label{eq:p_i^star}
\end{equation}
where $p_{r,i}^*$ is the equilibrium probability that a player with action rule $i$ ($i=1,2$) possesses reputation $r \in \{\rm{G,N,B}\}$, and
$\Phi_i^*$ represents the equilibrium probability that a $\sigma_i$ player receives $+$.
The equivalent of Eq.~\eqref{eq:p_i^star} and the following results can be obtained similarly for the saltatory reputation dynamics.
$\Phi_i^*$ is given by
\begin{equation}
 \Phi_i^*=\sum_{j=1}^2 \sum_{r \in \{\rm{G,N,B}\}} q_j p_{r,j}^* [\zeta_{r,i} \Phi_{{\rm C},r} + (1-\zeta_{r,i}) \Phi_{{\rm D},r}],
 \label{eq:Phi_i^star}
\end{equation}
where $\zeta_{r,i}$ is the probability that a $\sigma_i$ donor implements C when the recipient has reputation $r$.
Substitution of Eq.~\eqref{eq:p_i^star} into Eq.~\eqref{eq:Phi_i^star} leads to $\overrightarrow{\Phi^*} = g(\overrightarrow{\Phi^*})$, where $\overrightarrow{\Phi^*}=(\Phi_1^*,\Phi_2^*)$.

It should be noted that $\overrightarrow{\Phi^*} = g(\overrightarrow{\Phi^*})$ may have multiple fixed points.
An example is given by the population composed of the equal fraction of $\sigma_1={\rm gDisc}$ and $\sigma_2={\rm rDisc}$, i.e., $q_1=q_2=0.5$, under image scoring.
In this case, both a cooperative population (i.e., $\Phi_1^*,\Phi_2^* \approx (1-\epsilon_{\rm i})(1-\epsilon_{\rm a})$) and a defective population (i.e., $\Phi_1^*,\Phi_2^* \approx \epsilon_{\rm a}$) satisfy $\overrightarrow{\Phi^*} = g(\overrightarrow{\Phi^*})$.
Because of the multistability, we adopt $11^2=121$ initial conditions, i.e., $\overrightarrow{\Phi}=(0.1i,0.1j)$, $0 \le i,j \le 10$, for the iteration scheme to identify all the fixed points.
In fact, we find that the multistability does not cause a severe problem.
We will show in Results that five mixed populations composed of two action rules are stable and realize a sufficiently large probability of cooperation.
Among them, only one population, which consists of gDisc and rDisc and is stable under the so-called scoring-shunning social norm, yields the bistable equilibria.
One equilibrium yields a large cooperation probability ($>0.93$) and the other equilibrium yields a low cooperation probability ($<0.03$).
These results are qualitatively the same for the gradual and saltatory reputation dynamics.
For this social norm, we only keep the more cooperative equilibrium. 

Then, we obtain the trinary distribution of reputation for each of the two action rules by substituting $\Phi_i^*$ in Eq.~\eqref{eq:Phi_i^star}.
The average payoff per round to a $\sigma_i$ resident player ($i=1,2$) is given by
\begin{equation}
 \pi_i=-c \sum_{j=1}^2 q_j \Psi(\sigma_i,\{p_{{\rm G},j}^*,p_{{\rm N},j}^*,p_{{\rm B},j}^*\})+b \sum_{j=1}^2 q_j \Psi(\sigma_j,\{p_{{\rm G},i}^*,p_{{\rm N},i}^*,p_{{\rm B},i}^*\}),
 \label{eq:pi_i}
\end{equation}
where $\Psi(\sigma_i,\{p_{{\rm G},j}^*,p_{{\rm N},j}^*,p_{{\rm B},j}^*\})$ is the probability that a $\sigma_i$ donor cooperates with a $\sigma_j$ recipient.
In the equilibrium, the payoffs to the players with the different action rules are the same.
Therefore, we calculate the value of $b/c$ for which $\pi_1=\pi_2$.
If ${\rm d}(\pi_1-\pi_2)/{\rm d}q_1>0$, the mixed population is unstable against an infinitesimally small drift of the fraction of the two action rules.
We are concerned with the pairs of $\sigma_1$ and $\sigma_2$ that satisfy ${\rm d}(\pi_1-\pi_2)/{\rm d}q_1<0$ at the obtained $b/c$ value.

The mixed population is strict Nash when the payoff to any of the six mutants with a third action rule is smaller than that to a resident player.
By substituting Eq.~\eqref{eq:p^star2_mut} in 
\begin{equation}
 \Phi^{\prime *}=\sum_{j=1}^2 \sum_{r \in \{\rm{G,N,B}\}} q_j p_{r,j}^* [\zeta_r^\prime \Phi_{{\rm C},r} + (1-\zeta_r^\prime) \Phi_{{\rm D},r}],
 \label{eq:Phi^star_mut_hetero}
\end{equation}
we obtain the equilibrium probability that a mutant receives $+$ (i.e., $\Phi^{\prime *}$) and the equilibrium distribution of the mutant's reputation (i.e., $\{p_{\rm G}^{\prime *},p_{\rm N}^{\prime *},p_{\rm B}^{\prime *}\}$).
The payoff to a mutant player is given by
\begin{equation}
 \pi^\prime=-c \sum_{j=1}^2 q_j \Psi(\sigma^\prime,\{p_{{\rm G},j}^*,p_{{\rm N},j}^*,p_{{\rm B},j}^*\})+b \sum_{j=1}^2 q_j \Psi(\sigma_j,\{p_{\rm G}^{\prime *},p_{\rm N}^{\prime *},p_{\rm B}^{\prime *}\}).
 \label{eq:pi_mut2}
\end{equation}
As in the analysis of the homogeneous population, we also examine the stability of the heterogeneous population against invasion by the Self mutant (Sec.~\ref{sec:homo}).

We do not examine the mixed population composed of more than two action rules.
\section{Results}
We refer to each action rule by concatenating three letters, either C or D.
The first, second, and the third letters represent the intended action toward a recipient with reputation ${\rm G}$, ${\rm N}$, and ${\rm B}$, respectively.
For example, ${\rm gDisc}={\rm CCD}$ and ${\rm rDisc}={\rm CDD}$.
\subsection{Gradual Reputation Dynamics}\label{sec:gradual}
\subsubsection{Enumeration of Stable Populations}\label{sec:enumeration}
We set $\epsilon_{\rm i}=\epsilon_{\rm a}=0.02$ and consider the gradual reputation dynamics (Fig.~\ref{fig:dynamics}(A)).
We found that the homogeneous population is stable against invasion by mutants for some benefit-to-cost values $b/c$ for $108$ out of $8\times64=512$ action--norm pairs.
We exclude $64$ pairs with $\sigma={\rm AllD}$ from the $108$ pairs because of the lack of cooperation.
The entire game is symmetric with respect to the simultaneous flipping of ${\rm G} \leftrightarrow {\rm B}$ and $+ \leftrightarrow -$ \cite{OhtsukiIwasa2004JTB}.
Therefore, there are $(108-64)/2=22$ essentially distinct pairs.
The $22$ pairs are listed in Table~\ref{table:tab1}.
In addition, there are nine essentially distinct mixtures of two action rules that are stable under a certain social norm.
Among these stable populations, there are $12$ homogeneous populations composed of a single action rule and five heterogeneous populations composed of two action rules that realize a probability of cooperation larger than $0.5$ for a $b/c$ value smaller than $20$.
Our additional numerical simulations suggest that the probability of cooperation tends to unity if and only if this criterion is satisfied (Appendix B).

The cooperative equilibria, i.e., stable populations satisfying this criterion under a given social norm, are summarized in Fig.~\ref{fig:dynamicsA}(A).
In Fig.~\ref{fig:dynamicsA}(A), a social norm is represented by a combination of $s_{\rm G}$, $s_{\rm N}$, and $s_{\rm B}$, each of which takes either $+$$+$, $-$$+$, $-$$-$, or $+$$-$.
For example, $s_{\rm G}=-+$ indicates that donor's implemented action C and D toward a ${\rm G}$ recipient is assessed to be $-$ and $+$, respectively.
In fact, all the cooperative equilibria require $s_{\rm G}=+-$ such that only $s_{\rm N}$ and $s_{\rm B}$ are indicated in Fig.~\ref{fig:dynamicsA}(A).
\subsubsection{$s_{\rm N}=++{\rm \ or\ }-+$}\label{sec:*g}
Six action--norm pairs having $\sigma={\rm rDisc}$, $s_{\rm G}=+-$, $s_{\rm N}=++{\rm \ or\ }-+$, and $s_{\rm B}=++,\ -+,{\rm \ or\ }--$ realize a large probability of cooperation ($\approx 0.94$), a large probability of $+$ ($\approx 0.98$), and a mild restriction on $b/c$ (i.e., $b/c>1.004$).
In these equilibria, most resident players have the ${\rm G}$ reputation because the probability of $+$ is large.
A small fraction of players possesses the ${\rm N}$ reputation owing to error (see Sec.~\ref{sec:donation} for the definition of two types of error), and an even smaller fraction of players possesses the ${\rm B}$ reputation.
We verified that the population is stable against invasion by Self players (Sec.~\ref{sec:homo}) under each of the six social norms.

We call the six social norms standing--standing, standing--judging, standing--shunning, judging--standing, judging--judging, and judging--shunning.
The first (second) half of the name represents the social norm represented by $s_{\rm G}$ and $s_{\rm N}$ ($s_{\rm G}$ and $s_{\rm B}$) in the case of the binary reputation model.
For example, the combination of $s_{\rm G}=+-$ and $s_{\rm N}=++$ represents the standing social norm in the binary model if ${\rm N}$ is identified with ${\rm B}$ in the binary model (Fig.~\ref{fig:model}(D)).
Similarly, the combination of $s_{\rm G}=+-$ and $s_{\rm B}=-+$ represents the judging social norm in the binary model.
Therefore, we call the social norm given by $s_{\rm G}=+-$, $s_{\rm N}=++$, and $s_{\rm B}=-+$ standing--judging.

The six norms realize nearly perfect cooperation because the first half of the social norm (i.e., combination of $s_{\rm G}$ and $s_{\rm N}$) is either standing or judging and the rDisc donor cooperates with ${\rm G}$ recipients, but not ${\rm N}$ recipients.
It should be noted that standing and judging are the only second-order social norms that stabilize cooperation without special conditions in the binary reputation model \cite{OhtsukiIwasa2004JTB}.
Even if the donor's reputation transits from ${\rm N}$ to ${\rm B}$ owing to error, the donor regains the ${\rm N}$ reputation by cooperating with a ${\rm G}$ recipient, which occupies the majority of the population.
Although donors meeting recipients with the ${\rm B}$ reputation receive $-$ under $s_{\rm B}=--$, cooperation at the population level is maintained because few players have the ${\rm B}$ reputation.
\subsubsection{$s_{\rm N}=--$}\label{sec:bb}
Two cooperative action--norm pairs, i.e., $\sigma={\rm rDisc}$, $s_{\rm G}=+-$, $s_{\rm N}=--$, and $s_{\rm B}=++{\rm \ or\ }-+$, are also stable under a mild condition on $b/c$, i.e., $b/c>1.018$.
The homogeneous population of rDisc is not invaded by Self strategy under each of the two social norms.
However, the probability of $+$ ($\approx 0.74$) and that of cooperation ($\approx 0.66$) are not as large as those for the previous six action--norm pairs.
This is intuitively because, under the current social norms, i.e., shunning--standing and shunning--judging, a donor whose recipient has an ${\rm N}$ reputation always receives $-$ except in the case of the assessment error.
This behavior of the model is similar to that under shunning in the binary reputation model.
Nevertheless, different from the case of the binary model, the cooperation probability tends to unity in the error-free limit (Appendix B) for the two norms in the trinary reputation model.
We discuss this point further in Discussion.
\subsubsection{$s_{\rm N}=+-$}\label{sec:gb}
\begin{figure}[htbp]
 \begin{minipage}{0.49\linewidth}
  \begin{center}
  \includegraphics[width=\linewidth]{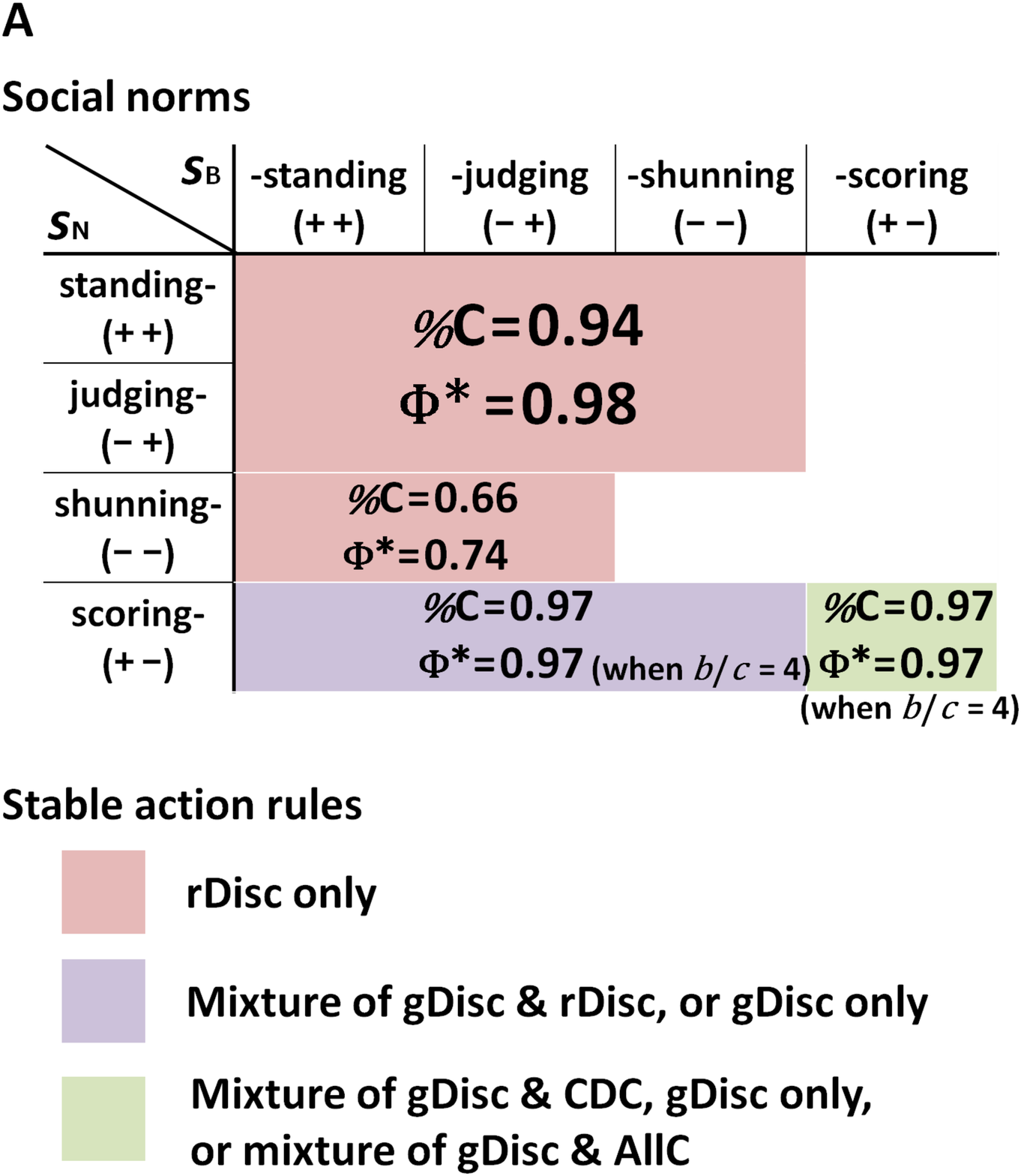}
  \end{center}
 \end{minipage}
 \begin{minipage}{0.49\linewidth}
  \begin{center}
  \includegraphics[width=\linewidth]{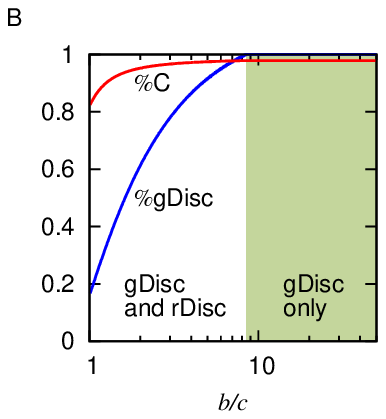}
  \includegraphics[width=\linewidth]{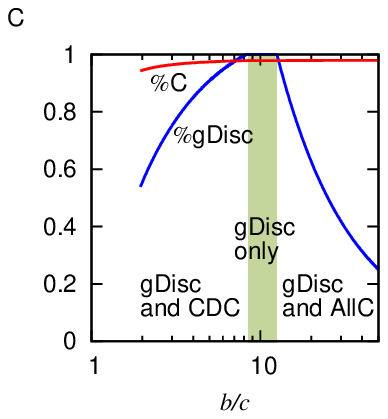}
  \end{center}
 \end{minipage}
 \caption{
Results for the gradual reputation dynamics.
We set $\epsilon_{\rm i}=\epsilon_{\rm a}=0.02$.
(A) Cooperative and stable action rules under different social norms.
All the shown social norms own $s_{\rm G}=+-$.
(B) Average cooperation probability and the fraction of gDisc players under scoring--standing (i.e., $s_{\rm G}=s_{\rm N}=+-$ and $s_{\rm B}=++$).
It should be noted that the fraction of gDisc and that of rDisc sum to unity.
(C) Average cooperation probability and the fraction of gDisc players under scoring--scoring (also called image scoring; $s_{\rm G}=s_{\rm N}=s_{\rm B}=+-$).
The fraction of gDisc and that of CDC or AllC sum to unity.
}
 \label{fig:dynamicsA}
\end{figure}

The other cooperative equilibria are four homogeneous populations and five heterogeneous populations, which are stable under social norms satisfying $s_{\rm G}=s_{\rm N}=+-$ (Fig.~\ref{fig:dynamicsA}(A)).
For $s_{\rm B}=++$ (scoring--standing) and $-$$+$ (scoring--judging), a homogeneous population composed of gDisc is stable for large $b/c$ ($>8.480$), and a mixed population composed of gDisc and rDisc is stable for small $b/c$ ($<8.480$).
For $s_{\rm B}=--$ (scoring--shunning), a homogeneous population composed of gDisc is stable for large $b/c$ ($>8.108$), and a mixed population composed of gDisc and rDisc is stable for small $b/c$ ($<8.108$).
The fraction of gDisc increases with $b/c$ under scoring--standing and is unity when $b/c>8.480$, as shown in Fig.~\ref{fig:dynamicsA}(B).
The results shown in Fig.~\ref{fig:dynamicsA}(B) are indistinguishable from those for scoring--judging and similar to those for scoring--shunning.
For the three social norms, the fraction of gDisc converges to $1-c/b$ in the limit $\epsilon_{\rm i},\epsilon_{\rm a} \rightarrow 0$ (Appendix C), which implies that only the mixed population of gDisc and rDisc is stable in the error-free limit.
Consistent with this, the threshold value of $b/c$ above which the homogeneous gDisc population is stable diverges as the error probabilities become small (Appendix B).

If we hypothetically merge ${\rm G}$ and ${\rm N}$, our result that cooperation is stable under scoring--standing and scoring--judging corresponds to the fact that the cooperation is stable under standing and judging, respectively, in the binary model \cite{OhtsukiIwasa2004JTB}.
Under scoring--shunning, cooperation is not undermined by $s_{\rm B}=--$ for the same reason as that for standing--shunning and judging--shunning (see Sec.~\ref{sec:*g} and Discussion).

Under scoring--scoring (i.e., $s_{\rm G}=s_{\rm N}=s_{\rm B}=+-$), which we also call the image scoring (Fig.~\ref{fig:model}(C)), there are three types of stable populations depending on the $b/c$ value.
When $1.941<b/c<8.230$, the mixed population composed of gDisc and CDC is stable (Fig.~\ref{fig:dynamicsA}(C)).
We regard CDC as a variant of rDisc because there are few players having the ${\rm B}$ reputation in the stable population; a CDC player obtains a slightly larger payoff than an rDisc player.
In fact, the mixed population of gDisc and rDisc is stable against invasion by all but CDC mutants.
When $8.230<b/c<12.53$, the homogeneous population of gDisc is stable (Fig.~\ref{fig:dynamicsA}(C)).
When $b/c>12.53$, the mixed population composed of gDisc and AllC is stable (Fig.~\ref{fig:dynamicsA}(C)).
In all the three cases, cooperation occurs with a large probability ($>0.94$).
This result is in a stark contrast with that in the binary reputation model, whereby image scoring does not usually support cooperation \cite{Panchanathan2003JTB,Ohtsuki2004JTB,OhtsukiIwasa2004JTB,Ohtsuki2007JTB,Roberts2008PRSB}.
We discuss this discrepancy in Discussion.

Under image scoring, the values of $b/c$ and the error probabilities under the assumption $\epsilon_{\rm i}=\epsilon_{\rm a}$, for which one of the three populations is stable are shown in Fig.~\ref{fig:scoring}(A).
For small error, the homogeneous population of gDisc and the mixed population composed of gDisc and AllC are invaded by rDisc mutants.
Only the heterogeneous population composed of gDisc and CDC survives the error-free limit.
In this limit, the fraction of gDisc is given by $1-c/b$ for the same reason as that for the heterogeneous population composed of gDisc and rDisc under scoring--standing (Sec.~\ref{sec:gb}; also see Appendix C).

The nine populations stable under $s_{\rm G}=s_{\rm N}=+-$ are unstable against invasion by Self mutants.
When there are more than two possible reputation values, Self is generally a strong competitor under image scoring such that it undermines cooperation \cite{Leimar2001PRSB}.
Our results are consistent with theirs.
In the next section, we propose a different reputation dynamics that makes cooperation stable against invasion by Self under image scoring.
\subsection{Saltatory Reputation Dynamics}\label{sec:other}
\subsubsection{Downward Saltation}\label{sec:down}
We first investigate a saltatory reputation dynamics (Fig.~\ref{fig:dynamics}(B)) given by $\beta_{\rm d}=0.5$ and $\beta_{\rm u}=0$.
This dynamics implies that a G reputation of a donor jumps down to a B reputation in one step with a probability $\beta_{\rm d}=0.5$.
We set $\epsilon_{\rm i}=\epsilon_{\rm a}=0.02$ and identify stable populations under each social norm.
To select cooperative equilibria, we imposed the same condition as that in the case of the gradual reputation dynamics, i.e., a cooperation probability larger than $0.5$ for some $b/c<20$.

We find that the set of cooperative equilibria (without consideration of Self mutants) is the same as that for the gradual reputation dynamics (Fig.~\ref{fig:dynamicsA}(A)).
The cooperation probability is equal to 
$0.9237$ for rDisc under standing--standing, 
$0.8614$ for rDisc under standing--shunning, 
$0.6804$ for rDisc under shunning--standing, 
$0.9340$ for gDisc under scoring--standing, and 
$0.9340$ for gDisc under scoring--shunning or image scoring.
The cooperation probability is preserved if we replace standing by judging.
Because a ${\rm G}$ reputation can turn into a ${\rm B}$ reputation in one step, the probability of cooperation is somewhat smaller than in the case of the gradual reputation dynamics.
The difference in the cooperation probability between the two reputation dynamics is relatively large when $s_{\rm B}$ is $-$$-$ or $+$$-$.
This is because players with the ${\rm B}$ reputation, albeit occupying a small fraction, can make other players with the ${\rm G}$ reputation to transit to the ${\rm B}$ reputation in one step under these social norms.

\begin{figure}[htbp]
 \begin{minipage}{0.32\linewidth}
  \begin{center}
  \includegraphics[width=\linewidth]{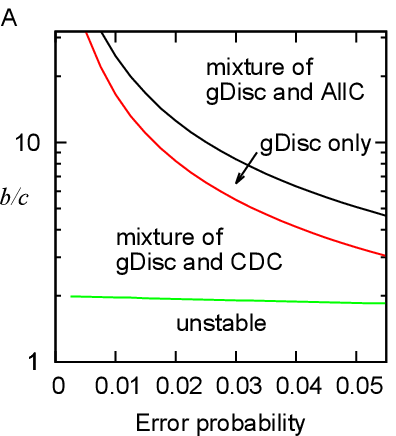}
  \end{center}
 \end{minipage}
 \begin{minipage}{0.32\linewidth}
  \begin{center}
  \includegraphics[width=\linewidth]{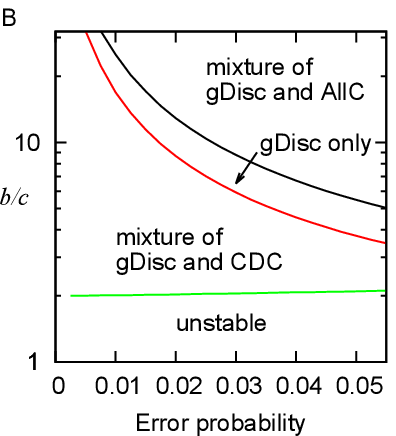}
  \end{center}
 \end{minipage}
 \begin{minipage}{0.32\linewidth}
  \begin{center}
  \includegraphics[width=\linewidth]{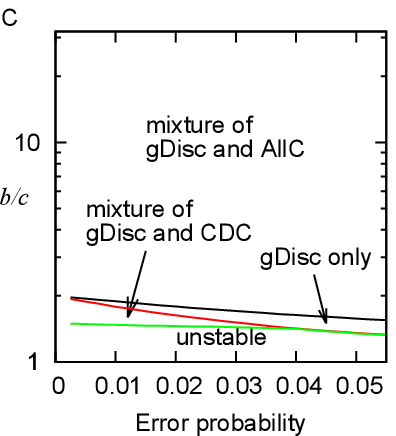}
  \end{center}
 \end{minipage}
 \caption{
Parameter regions in which one of the three populations is stable under image scoring.
We set the error probability $\epsilon_{\rm i}=\epsilon_{\rm a}$.
(A) Gradual reputation dynamics, i.e.,  $(\beta_{\rm d},\beta_{\rm u})=(0,0)$.
(B) Saltatory reputation dynamics with downward saltation, i.e., $(\beta_{\rm d},\beta_{\rm u})=(0.5,0)$.
(C) Saltatory reputation dynamics with upward saltation, i.e., $(\beta_{\rm d},\beta_{\rm u})=(0,0.5)$.
The cooperative population is stable against invasion by Self mutants in the parameter region above the green line in (B).
Self invades the cooperative population in all the parameter regions in (A) and (C).
}
 \label{fig:scoring}
\end{figure}

Under image scoring, the stable population as a function of $b/c$ and the error probabilities under the assumption $\epsilon_{\rm i}=\epsilon_{\rm a}$ is shown in Fig.~\ref{fig:scoring}(B).
Figure~\ref{fig:scoring}(B) is similar to the results for the gradual reputation dynamics shown in Fig.~\ref{fig:scoring}(A).
As a slight difference between Figs.~\ref{fig:scoring}(A) and \ref{fig:scoring}(B), less cooperative populations (e.g., a homogeneous population of gDisc as compared to a heterogeneous population of gDisc and AllC) can be stable in the downward saltatory reputation dynamics than in the gradual reputation dynamics for the same values of $b/c$ and the error probability.
This is intuitively because the downward saltatory reputation dynamics yields worse reputations than the gradual reputation dynamics.

We find that Self mutants cannot invade any of the stable populations shown in Fig.~\ref{fig:dynamicsA}(A).
\subsubsection{Upward Saltation}\label{sec:up}
We next identified the stable populations under the saltatory reputation dynamics (Fig.~\ref{fig:dynamics}(B)) given by $\beta_{\rm d}=0$, and $\beta_{\rm u}=0.5$.
This dynamics implies that a B reputation of a donor jumps up to a G reputation in one step with a probability $\beta_{\rm u}=0.5$.
We set $\epsilon_{\rm i}=\epsilon_{\rm a}=0.02$ and identify stable populations under each social norm.
To select cooperative equilibria, we imposed the same condition as that in the case of the gradual reputation dynamics, i.e., a cooperation probability larger than $0.5$ for some $b/c<20$.

We find that the set of cooperative equilibria (without consideration of Self mutants) is the same as that for the gradual reputation dynamics.
The cooperation probabilities in stable and cooperative populations are equal to 
$0.9416$ for rDisc under standing--standing, 
$0.9401$ for rDisc under standing--shunning, 
$0.7605$ for rDisc under shunning--standing, 
$0.9785$ for gDisc under scoring--standing, and 
$0.9783$ for gDisc under scoring--shunning or image scoring.
The cooperation probability is preserved if we replace standing by judging.
Because a ${\rm B}$ reputation can turn into a ${\rm G}$ reputation in one step, the probability of cooperation is a little larger than in the case of the gradual reputation dynamics.
The difference is relatively large when $s_{\rm N}=--$.

Under image scoring, the stable population as a function of $b/c$ and $\epsilon_{\rm i}$ $(=\epsilon_{\rm a})$ is shown in Fig.~\ref{fig:scoring}(C).
Figure~\ref{fig:scoring}(C) is quantitatively different from the results for the gradual reputation dynamics shown in Fig.~\ref{fig:scoring}(A).
More cooperative action rules (e.g., a heterogeneous population of gDisc and AllC compared to a homogeneous population of gDisc) are stable in the upward saltatory reputation dynamics than in the gradual reputation dynamics for the same values of $b/c$ and the error probability.
This is intuitively because the upward saltatory reputation dynamics yields better reputations than the gradual reputation dynamics.

We find that Self mutants cannot invade a homogeneous population of rDisc under each of the eight social norms satisfying $s_{\rm G}=+-$ and $s_{\rm N}=++,\ -+,\ --$.
However, under social norms satisfying $s_{\rm G}=s_{\rm N}=+-$, with image scoring included, the nine populations are not stable against invasion by Self mutants.
The result that Self undermines cooperation is the same as that for the gradual reputation dynamics (Fig.~\ref{fig:scoring}(A)).
\section{Discussion}
We analyzed a trinary reputation model of indirect reciprocity and identified cooperative Nash equilibria composed of a single action rule or mixture of two action rules.
Independent of details of the reputation dynamics (i.e., gradual or saltatory), we found at a small error level (i.e., $\epsilon_{\rm i}=\epsilon_{\rm a}=0.02$) that $12$ homogeneous populations and five mixed populations are cooperative and stable under different social norms (Fig.~\ref{fig:dynamicsA}(A)).
When the error probabilities are even smaller, eight homogeneous populations and four mixed populations remain stable (Appendix B).
In particular, under image scoring, the heterogeneous population composed of gDisc and CDC (a variant of rDisc) is stable in the error-free limit (Fig.~\ref{fig:scoring}(A); see Fig.~\ref{fig:model}(B) for the definition of gDisc, CDC, and rDisc).

The results derived from the trinary reputation model and those derived from the binary reputation model \cite{Nowak1998aJTB} are similar in some aspects.
For example, the standing--standing social norm (i.e., $s_{\rm G}=+-$, $s_{\rm N}=s_{\rm B}=++$) in the trinary model coincides with the standing social norm in the binary model (Fig.~\ref{fig:model}(D)) if we merge the ${\rm N}$ and ${\rm B}$ reputations in the trinary model.
Therefore, the result that standing--standing enables cooperation is not surprising.
Similarly, scoring--judging (i.e., $s_{\rm G}=s_{\rm N}=+-$, $s_{\rm B}=-+$), for example, is almost equivalent to judging in the binary model if we merge ${\rm G}$ and ${\rm N}$ reputations in the trinary model.
However, the results for the trinary and binary models are fundamentally different in the following two aspects.

First, shunning (see Fig.~\ref{fig:model}(D) in the case of the binary reputation model) is more supportive of cooperation in the trinary than binary model.
In the binary model, shunning results in a cooperation probability of $\approx 1/2$ in the error-free limit unless a cooperation prone initial condition and a finite number of rounds are combined \cite{Ohtsuki2007JTB} or reputations of players are only partially visible to others \cite{Nakamura2011PLos}.
In the trinary model, shunning--standing and shunning--judging stabilize full cooperation in the error-free limit (Appendix B).
Under these social norms, a donor meeting a recipient with an ${\rm N}$ reputation receives $-$ irrespective of the action, which is common to the behavior of the binary model.
In the trinary model, however, even if a donor obtains a ${\rm B}$ reputation, the donor easily receives $+$ either by cooperating with a ${\rm G}$ recipient or justifiably defecting against a ${\rm B}$ recipient.
Owing to the contribution of the justified defection, a donor gains $+$ more often than $-$ on average.
If players with the ${\rm G}$ reputation increase in number owing to this mechanism, players more likely receive $+$ than $-$.
This positive feedback sustains cooperation under shunning--related social norms in the trinary model.

Second, and the more important, image scoring is capable of supporting cooperation in our model.
Even if we consider Self mutants, which avoid a B reputation and are as selfish as possible, image scoring supports cooperation if the probability that a G reputation transits to a B reputation in one step is positive.
In previous literature, the Self strategy is recognized as a strong competitor that spoils cooperation in the indirect reciprocity game with more than two possible reputation values \cite{Leimar2001PRSB}.
Our conclusion that image scoring can support cooperation is consistent with the results derived from behavioral experiments \cite{Wedekind2000Science,Milinski2001PRSB,Seinen2006EER} and those derived from numerical simulations \cite{Nowak1998bNature,Diekmann2005ESSA} but opposite to those derived from the binary reputation model \cite{Panchanathan2003JTB,Ohtsuki2004JTB,OhtsukiIwasa2004JTB,Ohtsuki2007JTB}.
We reached this conclusion by simply introducing a third reputation to the standard binary model of indirect reciprocity.
The cooperation under image scoring in our model does not require forced cooperation in the first round \cite{Nowak1998aJTB}, partial cooperation of defective players \cite{Fishman2003JTB}, binomially or Poisson distributed number of rounds \cite{Brandt2004JTB,Brandt2006JTB}, growth of social networks used for transmission of reputation \cite{Brandt2005PNAS}, or a small probability with which the donor's reputation is revealed to other players \cite{Uchida2010PRE}.

In our model, cooperation under image scoring occurs for the following intuitive reason.
Although the composition of the stable population depends on the benefit-to-cost ratio and the error probabilities, the main action rule present in the stable population is gDisc, which cooperates with ${\rm G}$ and ${\rm N}$ recipients and defects against ${\rm B}$ recipients.
In the equilibrium, most gDisc resident players possess the ${\rm G}$ reputation.
If some AllD mutants are present, a ${\rm B}$ player in the population is likely to be AllD and not gDisc.
With the trinary reputation, gDisc players justifiably defect (i.e., D against a ${\rm B}$ recipient) but do not selfishly defect (i.e., D against a ${\rm G}$ or ${\rm N}$ recipient).
A donor that has justifiably defected would receive an ${\rm N}$ reputation but not a ${\rm B}$ reputation because few recipients have the ${\rm B}$ reputation in the population.
Therefore, gDisc players would not obtain the ${\rm B}$ reputation.
gDisc players that happen to obtain the ${\rm N}$ reputation, when selected as donor, likely meet a recipient with a ${\rm G}$ reputation such that they regain the ${\rm G}$ reputation.
In contrast, AllD players would obtain the ${\rm B}$ reputation such that they are not helped by others.
This contrasts with the case of the binary reputation model, in which a discriminating donor that defects against whatever recipient immediately receives the worst reputation (i.e., ${\rm B}$) and then is defected by others.

The presence of downward saltation (i.e., transition from the G reputation to B reputation in one step) is a key to make a cooperative population stable against invasion by Self mutants.
To explain this point intuitively, let us consider a homogeneous population of gDisc and assume that the implementation and assessment errors are absent.
If downward saltation is absent in the reputation dynamics, Self mutants flip between G and N reputations by alternatingly cooperating and defecting.
By doing so, the Self mutants can elicit cooperation from gDisc residents and cooperate with probability $\approx 1/2$.
Because gDisc residents cooperate with probability $\approx 1$, the gDisc population is invaded by Self mutants.
However, if downward saltation can occur in the reputation dynamics, a nonnegligible fraction of Self mutants possess the B reputation because they defect when they have the G reputation.
In contrast, gDisc players maintain the G reputation because they cooperate even if they have the G reputation.
Therefore, in the presence of downward saltation, Self mutants cannot invade the population of gDisc residents.

The mechanism of cooperation under image scoring in our model is similar to that under the so-called tolerant scoring proposed by Berger \cite{Berger2011GEB}.
The population of tolerant discriminator, which defects against a recipient if the recipient has defected in the last two rounds and cooperates otherwise, is stable.
Berger's results and ours are different in the following aspects.
First, Berger assumed three action rules, i.e., tolerant discriminator, which is similar to gDisc, AllC, and AllD and investigated the behavior of the model under tolerant scoring.
We carried out an exhaustive search of the space of the action rule and social norms to find the viability of image scoring.
Second, in our model, cooperation is stable even if the reputation moves only one step in a round, if the Self mutants are not considered.
It should be noted that Berger did not consider the Self mutants.
Third, cooperation is realized by a homogeneous population of one type of players in the Berger's model.
In our model, cooperation is realized by mixture of two types of discriminative players (i.e., gDisc and CDC).

The present study has following limitations.
First, we assumed that all the players in the population use the same social norm.
This oversimplification excludes a possible situation in which different norms compete in a population (e.g., \cite{Pacheco2006PLoS,Uchida2010PRE}).
Second, we only analyzed the stability of populations composed of up to two action rules for simplicity.
Third, we analyzed local stability of the equilibria and disregarded dynamics.
Even when a cooperative equilibrium is locally stable, it may in fact attract a tiny fraction of initial conditions.
Fourth, we assumed that the reputation sharing is public.
In other words, the information about the donor's new reputation immediately spreads from the observer to the entire population.
In a large population, such immediate spreading is impossible, and one has to assume, for example, that only a fraction of players gains the information about a donor in one game \cite{Nowak1998bNature,Brandt2004JTB,Uchida2010PRE}.

\section*{Appendix A: Distribution of the Reputation for the Saltatory Reputation Dynamics}
Under the saltatory reputation dynamics (Fig.~\ref{fig:dynamics}(B)), we obtain
\begin{eqnarray}
 \left \{
 \begin{array}{rcl}
  p_{\rm G}^*&=&p_{\rm G}^* \Phi^* + p_{\rm N}^* \Phi^* + p_{\rm B}^* \beta_{\rm u} \Phi^*,\vspace{-2mm}\cr
  p_{\rm N}^*&=&p_{\rm G}^* (1-\beta_{\rm d})(1-\Phi^*) + p_{\rm B}^* (1-\beta_{\rm u})\Phi^*,\vspace{-2mm}\cr
  p_{\rm B}^*&=&p_{\rm G}^* \beta_{\rm d}(1-\Phi^*) + p_{\rm N}^* (1-\Phi^*) + p_{\rm B}^* (1-\Phi^*).
 \end{array}
 \right .
\label{eq:ap_c}
\end{eqnarray}
Equation~\eqref{eq:ap_c} and the normalization $p_{\rm G}^*+p_{\rm N}^*+p_{\rm B}^*=1$ lead to
\begin{equation}
 \left [
 \begin{array}{c}
  p_{\rm G}^* \vspace{-2mm}\cr
  p_{\rm N}^* \vspace{-2mm}\cr
  p_{\rm B}^*
 \end{array}
 \right ] = \frac{1}{Z}\left [
 \begin{array}{c}
  (1-\beta_{\rm u}){\Phi^*}^2+\beta_{\rm u} \Phi^* \vspace{-2mm}\cr
  (1-\beta_{\rm d} \beta_{\rm u})\Phi^*(1-\Phi^*) \vspace{-2mm}\cr
  (1-\beta_{\rm d})(1-\Phi^*)^2+\beta_{\rm d}(1-\Phi^*)
 \end{array}
 \right ],
 \label{eq:ap_c2}
\end{equation}
where $Z=1-(1-\beta_{\rm d})(1-\beta_{\rm u})\Phi^*(1-\Phi^*)$.
\section*{Appendix B: Justification of the Criterion of Equilibrium Selection}
\begin{figure}[htbp]
  \begin{center}
  \includegraphics[width=\linewidth]{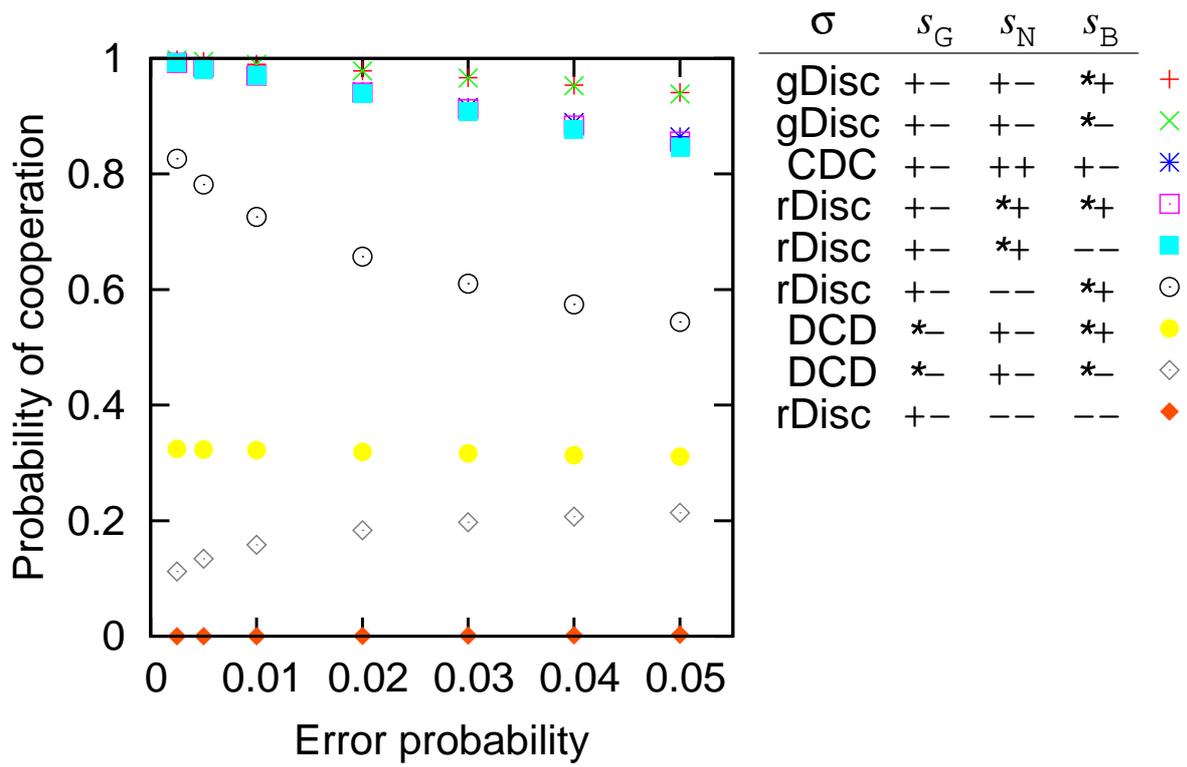}
  \end{center}
 \caption{
Relationships between the probability of cooperation and the error probability for the $22$ stable homogeneous populations.
We assume the gradual reputation dynamics and set $\epsilon_{\rm i}=\epsilon_{\rm a}$.
An asterisk represents either $+$ or $-$.
}
 \label{fig:robustness}
\end{figure}

We defined cooperative equilibrium as stable population in which the probability of cooperation is larger than $0.5$ for some $b/c$ such that $b/c<20$.
To justify this criterion, we carry out additional numerical simulations with various error probabilities satisfying $\epsilon_{\rm i}=\epsilon_{\rm a}$.
The probability of cooperation in each stable homogeneous population is shown for various error probability values in Fig.~\ref{fig:robustness}.
Figure~\ref{fig:robustness} shows that, under $13$ out of the $22$ stable action--norm pairs, the probability of cooperation seems to converge to unity in the error-free limit.
It should be noted that we identified $12$, not $13$, cooperative homogeneous populations in the main text.
One of the $13$ action--norm pairs is excluded because it is stable only for large $b/c$ values (i.e., $b/c>52.63$) at $\epsilon_{\rm_i}=\epsilon_{\rm_a}=0.02$ (Table~\ref{table:tab1}).
The probability of cooperation is larger than $0.5$ for some $b/c$ for five out of the nine stable heterogeneous populations composed of two action rules.
For all the five heterogeneous populations, the probability of cooperation is larger than $0.5$ for a $b/c$ value smaller than $20$ at $\epsilon_{\rm i}=\epsilon_{\rm a}=0.02$.

In the homogeneous populations, under eight out of the $13$ social norms for which the probability of cooperation seems to converge to unity in the error-free limit, i.e., standing--standing, standing--judging, standing--shunning, judging--standing, judging--judging, judging--shunning, shunning--standing, and shunning--judging, the range of $b/c$ in which the corresponding action rule is stable tends to $b/c>1$ as the error probabilities become small.
For example, under standing--standing, standing--judging, judging--standing, and judging--judging, the homogeneous population of rDisc is stable in the range $b/c>1.00006$ when $\epsilon_{\rm i}=\epsilon_{\rm a}=0.0025$; the corresponding range is $b/c>1.004$ when $\epsilon_{\rm i}=\epsilon_{\rm a}=0.02$ (Table~\ref{table:tab1}).
Under the other five social norms, i.e., standing--scoring, scoring--standing, scoring--judging, scoring--shunning, and scoring--scoring, stable cooperation requires a large value of $b/c$ when the error probabilities are small.
For example, under standing--scoring, the homogeneous population of CDC is stable in the range $b/c>402.1$ when $\epsilon_{\rm i}=\epsilon_{\rm a}=0.0025$; the corresponding range is $b/c>52.63$ when $\epsilon_{\rm i}=\epsilon_{\rm a}=0.02$ (Table~\ref{table:tab1}).
Under image scoring, the homogeneous population of gDisc is stable in the range $b/c>66.51$ when $\epsilon_{\rm i}=\epsilon_{\rm a}=0.0025$; the corresponding range is $8.230<b/c<12.53$ when $\epsilon_{\rm i}=\epsilon_{\rm a}=0.02$ (Table~\ref{table:tab1}, Fig.~\ref{fig:scoring}(A)).

In the heterogeneous populations, four out of five equilibria, i.e., mixture of gDisc and rDisc under scoring--standing, mixture of gDisc and rDisc under scoring--judging, mixture of gDisc and rDisc under scoring--shunning, and mixture of gDisc and CDC under image scoring, realize a large probability of cooperation in a wide range of $b/c$ when $\epsilon_{\rm i}=\epsilon_{\rm a}=0.02$ (Figs.~\ref{fig:dynamicsA}(B) and \ref{fig:dynamicsA}(C)) and also do so when the error probabilities are smaller.
For example, the heterogeneous population composed of gDisc and CDC is stable under image scoring in the range $1.994<b/c<66.51$ when $\epsilon_{\rm i}=\epsilon_{\rm a}=0.0025$; the corresponding range is $1.941<b/c<8.230$ when $\epsilon_{\rm i}=\epsilon_{\rm a}=0.02$ (Fig.~\ref{fig:scoring}(A)).
\section*{Appendix C: Fraction of Two Action Rules in the Error-free Limit}
We consider a social norm given by $s_{\rm G}=s_{\rm N}=+-$ and $s_{\rm B}={\rm ++,\ -+,{\rm \ or\ }--}$.
In the limit $\epsilon_{\rm i},\epsilon_{\rm a} \rightarrow 0$, the fraction of gDisc, denoted by $q_{\rm gDisc}$, and that of rDisc converge to $1-c/b$ and $c/b$, respectively, for the following reason.
In the equilibrium, few players possess the ${\rm N}$ or ${\rm B}$ reputation.
The behavior of rDisc players and that of gDisc players differ only toward recipients with reputation ${\rm N}$.
An rDisc donor with a ${\rm G}$ reputation defects against a recipient with an ${\rm N}$ reputation, receives $-$, and transits to the ${\rm N}$ reputation.
If this rDisc player is selected as recipient, an rDisc donor defects and a gDisc donor cooperates.
If selected as donor, this rDisc player would cooperate because most recipients have the ${\rm G}$ reputation, receive $+$, and transit to ${\rm G}$.
Because the rDisc player with an ${\rm N}$ reputation is selected as recipient once on average before selected as donor, the expected payoff to the rDisc donor during this period is equal to $q_{\rm gDisc}b$ on average.
During the same period, a gDisc donor with a ${\rm G}$ reputation cooperates with a recipient with an ${\rm N}$ reputation, pays $c$, receives $+$, keeps a ${\rm G}$ reputation, and gains benefit $b$ when selected as recipient.
By equating the payoffs to rDisc and gDisc players, we obtain $q_{\rm gDisc}b=-c+b$, which leads to $q_{\rm gDisc}=1-c/b$.
\section*{Acknowledgements}
We thank Mitsuhiro Nakamura for critical reading of the manuscript and the guidance on the literature on image scoring.
N.M. acknowledges the support provided through Grants-in-Aid for Scientific Research (No. 23681033, and Innovative Areas ``Systems Molecular Ethology''(No. 20115009)) from MEXT, Japan. 
This research is also supported by the Aihara Project, the FIRST program from JSPS, initiated by CSTP.

\clearpage
\begin{table}[htbp]
 \centering
  \caption{
Stable action--norm pairs.
We also show the range of $b/c$ in which action rule $\sigma$ is stable, probability of C, and mean $+$ assessment (i.e., $\Phi^*$) under the gradual reputation dynamics.
An asterisk represents either $+$ or $-$.
We set $\epsilon_{\rm i}=\epsilon_{\rm a}=0.02$.
}
 \label{table:tab1}
 \begin{tabular}{|c|c|c|c||l|c|c|}
  \hline
    $\sigma$ 		& $s_{\rm G}$ 		& $s_{\rm N}$ 		& $s_{\rm B}$ &\ \ \ \ \ range of $b/c$ & \%C & $\Phi^*$\\
  \hline \hline
    \multirow{3}{*}{CCD}&\multirow{3}{*}{$+$$-$}				&\multirow{3}{*}{$+$$-$}				&\textasteriskcentered $+$	&$8.480<b/c$		&0.9784			&0.9800\\
  \cline{4-7}
    			&						&						&$-$$-$				&$8.108<b/c$		&\multirow{2}{*}{0.9783}&\multirow{2}{*}{0.9783}\\
  \cline{4-5}
    			&						&						&$+$$-$				&$8.230<b/c<12.53$	&			&\\
  \hline
    CDC			&$+$$-$						&$+$$+$						&$+$$-$				&$52.63<b/c$		&0.9424			&0.9808\\
  \hline
    \multirow{3}{*}{CDD}&\multirow{3}{*}{$+$$-$}				&\textasteriskcentered $+$			&\textasteriskcentered $+$	&$1.004<b/c$		&0.9409			&0.9808\\
  \cline{3-7}
    			&						&\textasteriskcentered $+$			&$-$$-$				&$1.004<b/c$		&0.9392			&0.9792\\
  \cline{3-7}
    			&						&$-$$-$						&\textasteriskcentered $+$	&$1.018<b/c$		&0.6571			&0.7445\\
  \hline
    \multirow{4}{*}{DCD}&\textasteriskcentered $-$			&\multirow{4}{*}{$+$$-$}				&\textasteriskcentered $+$	&$3.716<b/c$		&0.3191			&0.5688\\
  \cline{2-2}\cline{4-7}
    			&$+$$-$						&						&\textasteriskcentered $-$	&$1.137<b/c<1.296$	&\multirow{3}{*}{0.1833}&\multirow{3}{*}{0.1833}\\
  \cline{2-2}\cline{4-5}
    			&$-$$-$						&						&$-$$-$				&$1.120<b/c$		&			&\\
  \cline{2-2}\cline{4-5}
    			&$-$$-$						&						&$+$$-$				&$1.120<b/c<11.66$	&			&\\
  \hline
    CDD			&$+$$-$						&$-$$-$						&$-$$-$				&$25.54<b/c$		&0.0004			&0.0004\\
  \hline
 \end{tabular}
\end{table}
\end{document}